\documentclass[useAMS,usenatbib]{mn2e}

\pdfoutput=1

\usepackage{graphicx,natbib,aas_macros,amssymb,amsmath,enumerate}
\usepackage{deluxetable}

\title[Neutron Star Kicks]{The Hydrodynamic Origin of Neutron Star Kicks}

\author[J.~Nordhaus et al.]{J.~Nordhaus$^{1,2,3,4}$\thanks{NSF Astronomy and Astrophysics Postdoctoral Fellow; E-mail: nordhaus@astro.rit.edu}, 
T.~D.~Brandt$^{4}$, 
A.~Burrows$^{4}$, 
A.~Almgren$^{5}$\\
$^1$ Center for Computational Relativity and Gravitation, Rochester Institute of Technology, Rochester, NY 14623, U.S.A.\\
$^2$ National Technical Institute for the Deaf, Rochester Institute of Technology, Rochester, NY 14623, U.S.A.\\
$^3$ Department of Physics and Astronomy, University of Rochester, Rochester, NY 14627, U. S. A.\\
$^4$ Department of Astrophysical Sciences, Princeton University, Princeton, NJ 08544, U.S.A.\\
$^5$ Computational Research Division, Lawrence Berkeley National Lab, Berkeley, CA 94720, U.S.A.}

\begin{document}
\date{Submitted XXX}
\pubyear{2011}
\maketitle
\label{firstpage}

\begin{abstract}
We present results from a suite of axisymmetric, core-collapse
supernova simulations in which hydrodynamic recoil from an asymmetric
explosion produces large proto-neutron star (PNS) velocities.  We use
the adaptive-mesh refinement code CASTRO to self-consistently follow
core-collapse, the formation of the PNS and its subsequent
acceleration.  We obtain recoil velocities of up to 620 km\,s$^{-1}$
at $\sim$1 s after bounce.  These velocities are consistent with the
observed distribution of pulsar kicks and with PNS velocities obtained
in other theoretical calculations.  Our PNSs are still accelerating at
several hundred km\,s$^{-1}$ at the end of our calculations,
suggesting that even the highest velocity pulsars may be explained by
hydrodynamic recoil in generic, core-collapse supernovae.
\end{abstract}


\section{Introduction}
At birth, pulsars achieve velocities well above those of their progenitor population \citep{Gunn:1970ys,Lyne:1994rr}.  These pulsar ``kicks" typically range from 200 km\,s$^{-1}$ to 400 $\rm km$\,$\rm s^{-1}$, with the fastest neutron stars achieving velocities near, or in excess of, 1000 km\,s$^{-1}$ \citep{Lyne:1994rr,Chatterjee:2005uq,Hobbs:2005fk}.  The current distribution of observed pulsar velocities is Maxwellian, hinting at a common acceleration mechanism \citep{Hansen:1997dq,Hobbs:2005fk,Zou:2005yq,Faucher-Giguere:2006lr}.

Many scenarios have been proposed for the origin of pulsar recoil and
neutron star kicks.  Popular mechanisms often require strongly
magnetized systems, exotic neutrino physics, and/or rapid rotation to
produce substantial kicks.  For example, in the presence of strong
magnetic fields, neutrino-matter interactions can generate neutron
star velocities on the order of a few hundred km\,s$^{-1}$ by
producing $\sim$1\% dipole asymmetries
\citep{Lai:1998tg,Nardi:2001hc,Lai:2001ij,Kusenko:1999kx,Lambiase:2005yq,Barkovich:2004vn,Fuller:2003yq,Kishimoto:2011fk}.
Many of these scenarios require magnetic fields in the magnetar range
(i.e.~10$^{14-16}$ G) and may not produce substantial kicks in typical
core-collapse supernovae.  Other scenarios involve jet/counter-jet
misalignment launched near the proto-neutron star (PNS).  In such
situations, the jets accompany an associated gamma-ray burst (GRB) or
form through magneto-rotational processes during core collapse
\citep{Cen:1998oq,Khokhlov:1999qy,Sawai:2008kx,Papish:2011qy}.  These
scenarios require rapid rotation and therefore, may only manifest in a
small subset of core-collapse events.

If neutron star kicks are a generic outcome of core collapse, then a
natural explanation is recoil during an asymmetric supernova
explosion.  In the current, most sophisticated simulations, the bounce
shock, launched when the equation of state stiffens at nuclear
densities, stalls due to thermal energy losses from neutrino emission
and dissociation of nuclei into nucleons.  The stalled shock itself is
subject to hydrodynamic and neutrino-driven instabilities, which
manifest as prominent low-order $\ell$-mode oscillations in
axisymmetric simulations of non-rotating progenitors
\citep{Blondin:2003ul,Scheck:2004rt,Buras:2006qf,Scheck:2006vn,Burrows:2007eu,Blondin:2007mz,Ott:2008cr,Fernandez:2009uq,Nordhaus:2010fr,Nordhaus:2010kx,Fernandez:2010ly,Brandt:2011fj}.
At the onset of shock revival, the PNS may recoil if large-scale
asymmetries are present during the ensuing supernova explosion.  While
the mechanism by which core-collapse supernova progenitors explode is
not fully understood, the most probable scenario involves absorption
of neutrinos in the post-shock ``gain region" \citep{Bethe:1985fr} and
likely requires the development of multi-dimensional instabilities in
fully three-dimensional radiation-hydrodynamic simulations to succeed
\citep{Nordhaus:2010fr}.  Nonetheless, recoil from an asymmetric
neutrino-driven explosion presents a natural mechanism by which PNSs
achieve high velocities
\citep{Burrows:1996th,Scheck:2004rt,Scheck:2006vn,Nordhaus:2010kx,Wongwathanarat:2010yq}
and appears to be supported by recent X-ray observations of the
Cassiopeia A supernova remnant \citep{Hwang:2011lr}.

Computational studies of recoil require multi-dimensional, radiation-hydrodynamics calculations which start at the onset of collapse and follow the dynamics self-consistently.  This includes the formation of the PNS, the explosion, the propagation of the shock front through the stellar envelope and eventually, decoupling of the PNS from the surrounding material.  Such an approach is computationally challenging and as such, various techniques have been adopted to make the computations tractable.  One popular approach is to commence the calculations after bounce by mapping spherically symmetric solutions onto a multi-dimensional grid and excising the PNS from the computational domain \citep{Scheck:2004rt,Scheck:2006vn,Wongwathanarat:2010yq}.  This requires one to infer a PNS kick through a rigid, impenetrable inner boundary, but allows one to track the supernova explosion for several seconds and to distances greater than 10,000 km.  While this approach is appealing, it must be checked by simulations which include the PNS in the computational domain.

Recently, we have carried out the first axisymmetric,
radiation-hydrodynamic simulation of recoil with the multi-group,
arbitrary, Lagrangian-Eulerian code {\sc VULCAN/2D}
\citep{Nordhaus:2010kx}.  By transitioning from a spherical-polar mesh
to a pseudo-Cartesian mesh at the center of the domain, we
self-consistently tracked the PNS's formation and off-center motion.
This calculation was computationally expensive, as it employed
multi-group flux limited diffusion neutrino transport and followed the
supernova explosion until the shock reached the 5,000 km radial outer
boundary of the domain.  At that time, the PNS had reached a velocity
of $\sim$150 km\,s$^{-1}$ but had yet to fully decouple from the
ejecta.  The PNS recoil was due almost entirely to hydrodynamical
processes and was consistent with previous excised-core calculations \citep{Scheck:2004rt,Scheck:2006vn}.  Asymmetric neutrino emission contributed $\sim$2\% to the
overall kick magnitude.  This suggests that neutrinos play no
significant role in accelerating neutron stars to high velocities
during typical core-collapse supernovae.  At the end of the
calculation, significant acceleration ($\sim$350 km\,s$^{-2}$), in
addition to the degree of asymmetry in the ejecta, suggested that
higher PNS velocities were possible.  Verifying these estimates
requires tracking the supernova shock to greater radial distances and
later times.

To expand upon the work of \cite{Nordhaus:2010kx}, we carry out a
suite of axisymmetric collapse calculations with the
adaptive-mesh-refinement (AMR), radiation-hydrodynamics code, CASTRO
\citep{Almgren:2010fk,Zhang:2011lr}.  By employing AMR and a
simplified form of transport (see Sec. \ref{2}), we expand the domain
to a radial distance of 10,000 km and perform multiple calculations.
Our PNSs achieve recoil velocities ranging from tens of km\,s$^{-1}$
up to $\sim$620 km\,s$^{-1}$.  In general, the magnitude of the recoil
depends on the degree of asymmetry at the time of explosion and the
energy of the explosion itself \citep{Burrows:2007ly}.  In
Sec.~3, we discuss the physical processes that accelerate the
PNS.  In Sec. \ref{4}, we compare our results with previous work
before concluding and discussing future work in Sec. \ref{5}.

\section{Computational Methodology\label{2}}
We carry out our simulations using the AMR, radiation-hydrodynamics
code, CASTRO \citep{Almgren:2010fk,Zhang:2011lr}.  CASTRO is a finite-volume, Eulerian code which employs an unsplit version of the piecewise parabolic method (PPM) for the hydrodynamics and a multigrid Poisson solver to handle self-gravity.  To facilitate multiple
calculations, we simplify our radiation transport by using radius- and
temperature-dependent prescriptions for the neutrino heating and
cooling rates $\mathcal{H}$ and $\mathcal{C}$.  We solve the fully
compressible equations of hydrodynamics:
\begin{align}
\partial_t\rho &= -\nabla\cdot\left(\rho \mathbf{v}\right) \label{eq:hydro}\\
\partial_t\left(\rho\mathbf{v}\right) &= -\nabla\cdot\left(\rho \mathbf{v}\mathbf{v}\right) 
- \nabla p + \rho \mathbf{g}\nonumber\\
\partial_t\left(\rho E\right) &=
-\nabla\cdot\left(\rho\mathbf{v}E+p\mathbf{v}\right)+
\rho\mathbf{v}\cdot\mathbf{g}+\rho(\mathcal{H}-\mathcal{C})\, , \nonumber
\end{align}
where $p$, $T$, $\rho$, $\mathbf{g}$, and $\mathbf{v}$ are the fluid
pressure, temperature, density, gravitational acceleration, and
velocity.  The specific total energy is given as $E = e + \frac{1}{2}v^2$,
where $e$ represents the internal energy.  Neutrino heating and
cooling occur via the super-allowed charged-current reactions
involving free nucleons, electrons, protons, electron neutrinos and anti-electron neutrinos.  We
use the heating and cooling rates derived in \cite{Janka:2001rr} and
previously used by \cite{Murphy:2008dq} and \cite{Nordhaus:2010fr}.
These rates, assuming the electron and anti-electron neutrino
luminosities, $L_{\nu_e}$ and $L_{\bar{\nu}_e}$, to be equal, are
\begin{align}
\mathcal{H} = 1.544&\times 10^{20} \left(\frac{L_{\nu_e}}{10^{52}\ 
{\rm erg}\ {\rm s}^{-1}}\right) \left(\frac{T_{\nu_e}}{4\ {\rm MeV}}\right)^2\times\label{eq:heating}\\
 &\left(\frac{100 {\rm km}}{r}\right)^2 \left(Y_{\rm n} + Y_{\rm p}\right) 
e^{-\tau_{\rm eff}} \left[\frac{{\rm erg}}{{\rm g}\ {\rm s}}\right]\nonumber
\end{align}

and

\begin{equation}
\mathcal{C} = 1.399\times 10^{20} \left(\frac{T}{2\ {\rm MeV}}\right)^6 
\left(Y_{\rm n} + Y_{\rm p}\right) e^{-\tau_{\rm eff}}
\left[\frac{{\rm erg}}{{\rm g}\ {\rm s}}\right]\label{eq:cooling}\, ,
\end{equation}
where $T_{\nu_e}$ is the electron neutrino temperature, $r$ is the
distance from the center of the star, $Y_{\rm n}$ and $Y_{\rm p}$ are
the neutron and proton fractions, and $\tau_{\rm eff}$ is an effective optical depth for electron
and anti-electron neutrinos (see Eqs. 6 \& 7 of \cite{Hanke:2011lr}).
The factor $e^{-\tau_{\rm eff}}$ effects a transition between the
dense inner regions, where neutrinos are trapped and heating and
cooling are suppressed, to the outer, optically thin regions.  To
avoid the prohibitive cost of global calculations, we fit $\tau_{\rm
eff}$ as a function of density by post-processing models and applying
Eq. 6 of \cite{Hanke:2011lr}.  In the transition region, $10^{-2}
\lesssim \tau_{\rm eff} \lesssim 1$, the resulting power-law fits give
root-mean-square residuals of $\sim$10\% in $e^{-\tau_{\rm eff}}$.  Note also that the bulk of neutrino heating occurs within the inner few hundred kilometers and falls off rapidly with radial distance.  The electron fraction, $Y_{\rm e}$, evolves on
infall via a density prescription given in \cite{Liebendorfer:2005lq}
and previously employed by \cite{Murphy:2008dq} and
\cite{Nordhaus:2010fr}.  To close the equations, we use the
sophisticated, finite temperature, nuclear equation of state of
\cite{Shen:1998gf}.

The current theoretical consensus holds that most 2D
core-collapse simulations do not produce neutrino-induced explosions
for the majority of progenitors without supplementing the neutrino
luminosity \citep{Burrows:2007ly,Ott:2008cr,Nordhaus:2010kx,Brandt:2011fj,Fujimoto:2011fj}.  However, explosions in axisymmetric simulations have been obtained after $\sim$600 ms for a 15 $M_\odot$ progenitor when a soft nuclear equation of state and variable Eddington factor closure technique are employed \citep{Marek:2009lr}.
We therefore induce explosions by varying the neutrino driving
luminosity $L_{\nu_e}$ ($= L_{\bar{\nu}_e}$) from 2.1 to $2.5 \times 10^{52}$ erg\,s$^{-1}$.
These values are comparable to the time-dependent values achieved in 2D
calculations with detailed neutrino transport
\citep{Ott:2008cr,Brandt:2011fj}.  Our simplified transport scheme couples
this energy somewhat more efficiently to the post-shock region, and
our results should be checked by future studies that employ sophisticated, but computationally expensive, neutrino transport.  We hold
the driving luminosity $L_{\nu_e}$ fixed in each calculation, but vary
it from run to run.  Our simplified transport method, in conjunction
with AMR, allows us to follow collapse, formation of the PNS,
and any subsequent acceleration due to the supernova explosion.

We use the 15-$M_\odot$, red-supergiant, non-rotating,
solar-metallicity progenitor of \cite{Woosley:1995mz}.  Our
simulations begin at the onset of core-collapse, continue through the
formation of the PNS and the launching, stalling, and revival of the
bounce shock, and end when the shock approaches the edge of the
computational domain.  We employ a 2D, 10,000-km by 20,000-km domain
discretized with a uniform coarse grid of 64 by 128 cells covering the
full 180$^{\circ}$ range in polar angle.  Within 200 km of the
PNS, we use 4 AMR levels, each increasing the resolution by a
factor of 4, giving a minimum cell size of $\sim$0.3 km.  Exterior to
200 km, the adaptive mesh tracks the high entropy, shocked material.

We index our eight simulations by their driving electron neutrino
luminosity $L_{\nu_e}$ in units of $10^{52}$ erg\,s$^{-1}$, from
$L_{2.1}$ to $L_{2.5}$.  For each simulation, we follow the explosion
for $\sim$1 second of post-bounce evolution.  For all but the
$L_{2.1}$ model, the calculation ends when the shock approaches the
edge of the computational domain, 10,000 km from the PNS.  At this
point the total momentum on the grid is conserved to within $\sim$1\%
of the core's final value.

\section{Acceleration of the PNS\label{3}}

The hydrodynamic flow behind the stalled shock is turbulent, and
is soon deformed by the development of low-mode hydrodynamic and
neutrino-driven instabilities
\citep{Burrows:1995zr,Blondin:2003ul,Scheck:2004rt,Scheck:2006vn,Blondin:2007mz,Fernandez:2010ly}.
This guarantees an asymmetric distribution of material when neutrino
heating revives the stalled shock.  The ensuing asymmetric explosion
accelerates the PNS as the shock propagates through the stellar
envelope.

\begin{figure}
\begin{center}
\includegraphics[width=8cm,angle=0,clip=true]{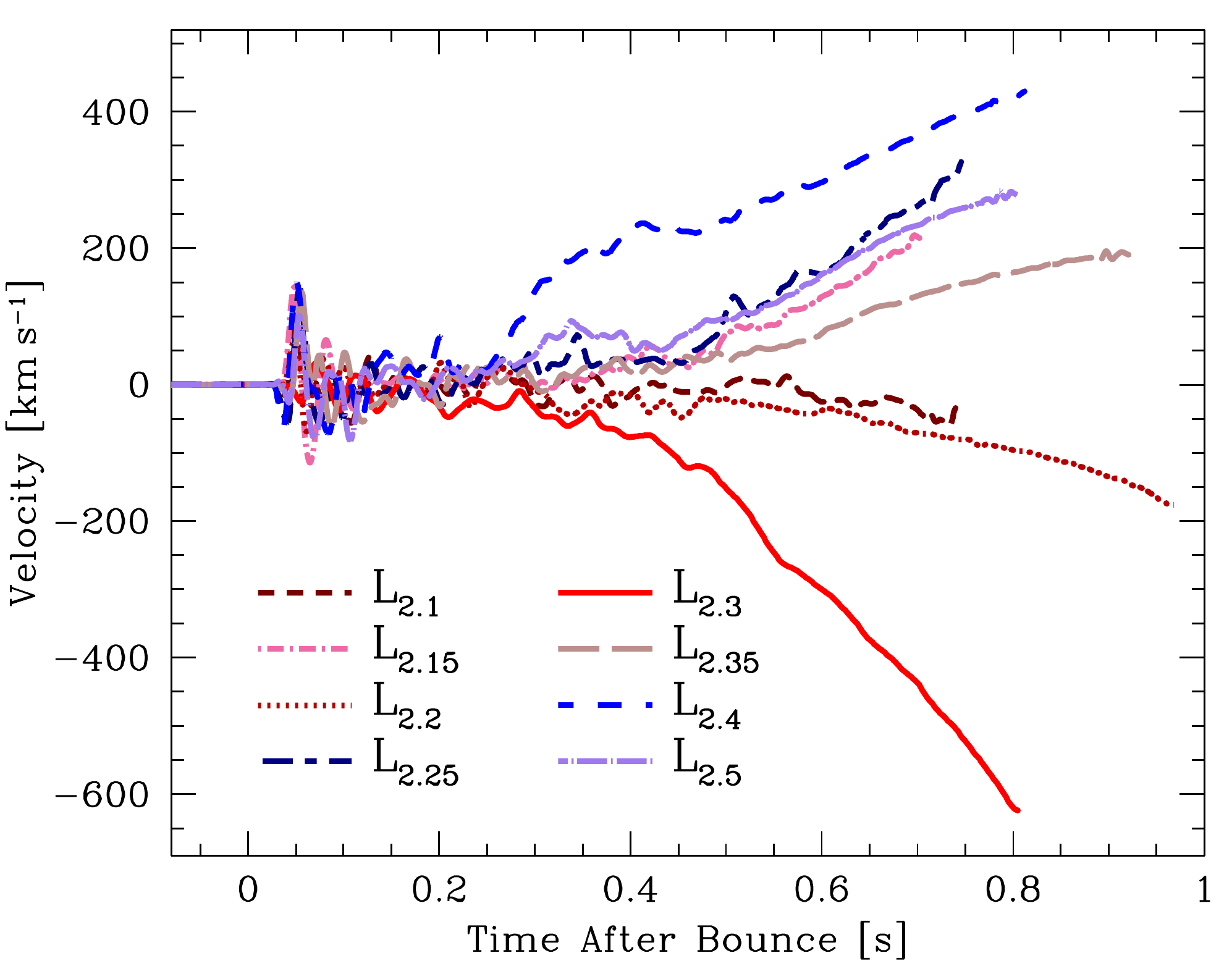}
\caption{PNS recoil velocities as a function of time after formation of the bounce shock.  The subscript in the simulation names refers to the electron neutrino, $\nu_e$, and anti-electron neutrino, $\bar{\nu}_e$, luminosities, each in units of $10^{52}$ erg s$^{-1}$ (see Eqs. \eqref{eq:heating} and \eqref{eq:cooling} and Table \ref{table1}).  
\label{fig:corevels}}
\end{center}
\end{figure}

To understand the physical processes governing the recoil, we first
present the PNS velocities obtained in our simulations.  We compute
the core positions as the centroids of the density distributions and
differentiate to obtain the velocities, which we show as functions of
time in Figure \ref{fig:corevels}.  The subscript in each simulation
label indicates the driving $L_{\nu_e}$ in units of $10^{52}$
erg\,s$^{-1}$.  The value of the driving $L_{\bar{\nu}_e}$ is taken to be the same, thus yielding a total driving luminosity in each simulation of $L_{\rm tot}=L_{\nu_e} + L_{\bar{\nu}_e} = 2 L_{\nu_e}$.  We obtain PNS velocities ranging from tens to many
hundreds of km\,s$^{-1}$.  The highest velocity PNS (solid
red curve in Fig.~\ref{fig:corevels}) reaches a speed of 624 km
s$^{-1}$ after $\sim$800 ms of post-bounce evolution, and is still
accelerating at $\sim$1000 km\,s$^{-2}$.

Figure \ref{fig:L_2p4_348} depicts the evolution of a representative run
(model $\rm L_{2.4}$).  The left column presents the evolution of
electron fraction, $Y_{\rm e}$, over the inner $\sim$200 km, as the
explosion progresses.  The middle of the computational domain is
marked by a black line ($Z=0$).  The PNS is clearly visible as the
deleptonized blue region, which begins at $Z=0$ and moves off-center.
The right column shows the global evolution of the anisotropic
supernova shock, with the color map depicting entropy.  The explosion
occurs primarily in the $-Z$ direction while the PNS recoils
in the $+Z$ direction.  By $\sim$800 ms after bounce, the PNS has
largely decoupled from its surroundings.  The middle column shows the density distribution with the shock outlined in pink.  A high-density region above the PNS combines with a low-density region below it to gravitationally accelerate the PNS in the $+Z$ direction.

\subsection{\label{subsec:interp}Physics of the Recoil\protect}

To determine the physical processes governing the movement of the PNS,
we post-process our results by computing the hydrodynamic acceleration
$\vec{a}_{\rm c}$ of the core due to anisotropic pressure forces,
momentum flux, and gravitational forces.  The Eulerian equations of
hydrodynamics give
\begin{equation}
\vec{a}_{\rm c}=\dot{\vec{v}}_{\rm c}\approx\int_{r> r_{\rm
    c}}\frac{G\vec{r}}{r^3}dm - \frac{1}{M_{\rm
    c}}\left[\oint_{r=r_{\rm c}} P d\vec{S} + \oint_{r=r_{\rm c}}\rho v_{\rm r}\vec{v}dS \right],
\label{eq:acceleration}
\end{equation}
where $\rho$ is the density, $M_{\rm c}$ and $\vec{v}_{\rm c}$ are the
mass and mean velocity of the core, $P$ is the gas pressure, $\vec{v}$
is the fluid velocity, $v_{\rm r}$ is the radial component of the
velocity, and $r_\mathrm{c}$ is a fiducial spherical radius
\citep{Scheck:2006vn,Nordhaus:2010kx}.

The three terms in Eq.~\eqref{eq:acceleration} represent the
contributions to the acceleration from the gravitational field of
matter exterior to $r_{\rm c}$, anisotropic gas pressure, and momentum flux
through $r_{\rm c}$, respectively.  The first term assumes a
spherically-symmetric matter distribution interior to $r_{\rm c}$,
which in practice is an excellent approximation.  In the limit of an
isotropic explosion, no acceleration occurs and each term vanishes
individually.


\begin{figure}
\begin{minipage}{180mm}
\begin{onecolumn}
\begin{center}
\includegraphics[width=9.5cm,angle=0,clip=true]{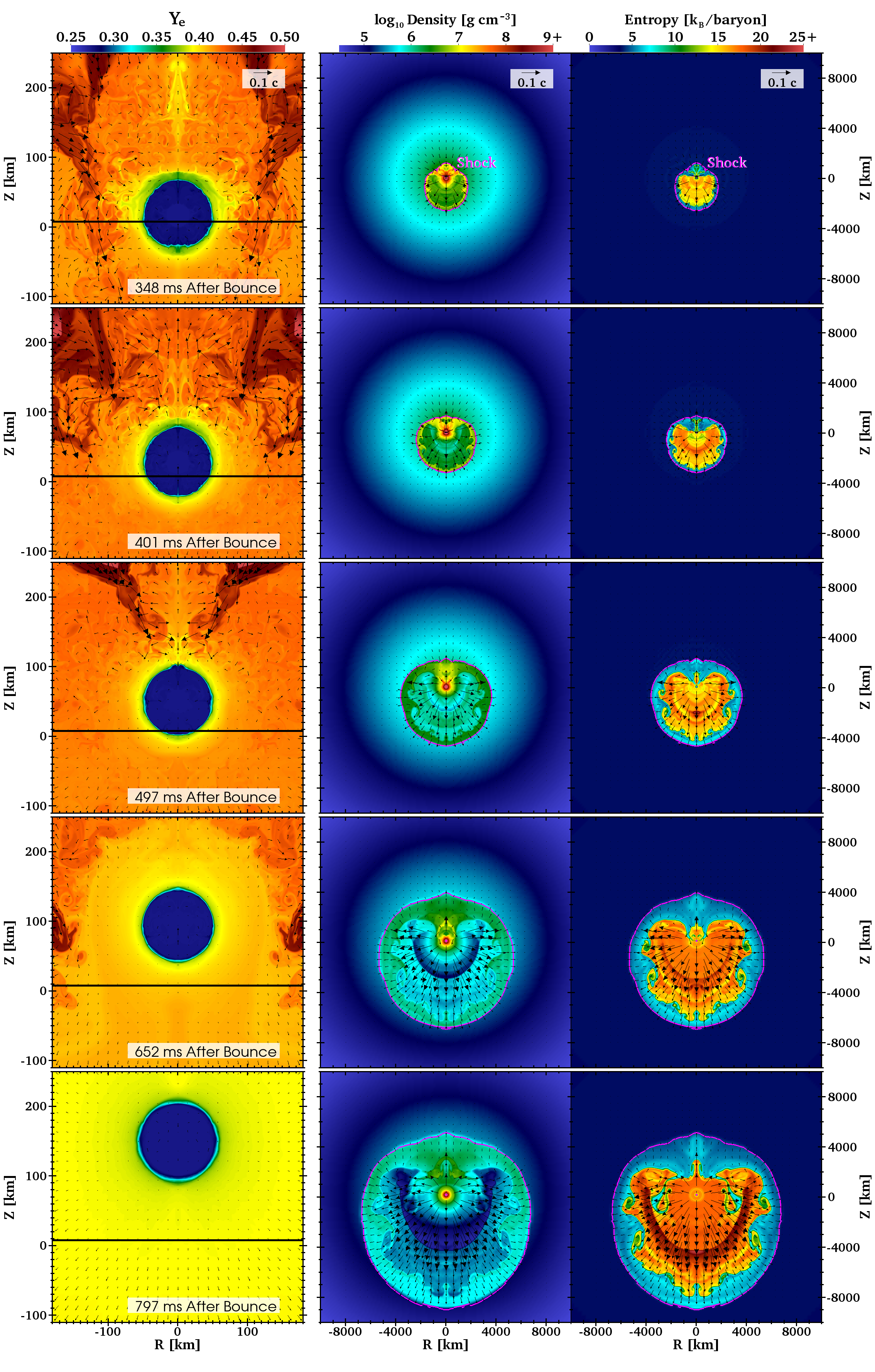}
\caption{The left column shows the evolution of the inner $\sim$200 km of the
  simulation domain for model $\rm L_{2.4}$.  The color map depicts the electron fraction,
  $Y_{\rm e}$, with velocity vectors overlaid; the black lines
  indicate $Z=0$.  The color map for the middle column depicts density.  The pink curve shows the location of the shock.  Exterior to the shock, the flow is radially inward.  The right column panels show the entropy evolution of the
  explosion.  The steep entropy jump just interior to the shock depicts the region where nucleons are reassociated into nuclei and alpha particles.  At 707 ms after bounce, the PNS wind is seen as the dark blue region interior to the shock in the middle panel.  The supernova explosion primarily occurs in the $-Z$ direction,
  while the PNS recoils in the $+Z$ direction.  At $\sim$800 ms after bounce, the PNS has largely decoupled from the surrounding material,
  but is still being accelerated by the gravitational pull of
  slow-moving ejecta in the $+Z$ direction (see
  Fig.~\ref{fig:accel_3panel}).
\label{fig:L_2p4_348}}
\end{center}
\end{onecolumn}
\end{minipage}
\end{figure}

\begin{twocolumn}

Equation \eqref{eq:acceleration} includes contributions from
hydrodynamic processes, but neglects radiation pressure asymmetries,
which are not captured by our heating and cooling prescription
(Eqs.~\eqref{eq:heating}-\eqref{eq:cooling}).  In our previous
calculations, which performed radiative transfer using multi-group
flux-limited diffusion, neutrino momentum contributed $\lesssim2$\% of
the overall kick \citep{Nordhaus:2010kx}.  This is consistent with
previous studies that found neutrino radiation pressure to contribute
$\sim$5\% to the final kick velocity
\citep{Scheck:2004rt,Scheck:2006vn}.

The relative contributions of the terms in Eq.~\eqref{eq:acceleration}
depend on the properties of the flow and the explosion dynamics.  As a
consequence of the explosion, the PNS generically recoils away from
the high-velocity ejecta and towards the lower-velocity ejecta.
However, the interpretation of the kick is not as straightforward as
Eq.~\eqref{eq:acceleration} would suggest.  As discussed in Section III
of \cite{Nordhaus:2010kx}, pressure and gravity do work on fluid
elements; anisotropic pressure or gravitational forces at a small value of
$r_{\rm c}$ will appear as anisotropic momentum flux at a larger value
of $r_{\rm c}$.

\begin{figure}
\begin{center}
\includegraphics[width=7.5cm,angle=0,clip=true]{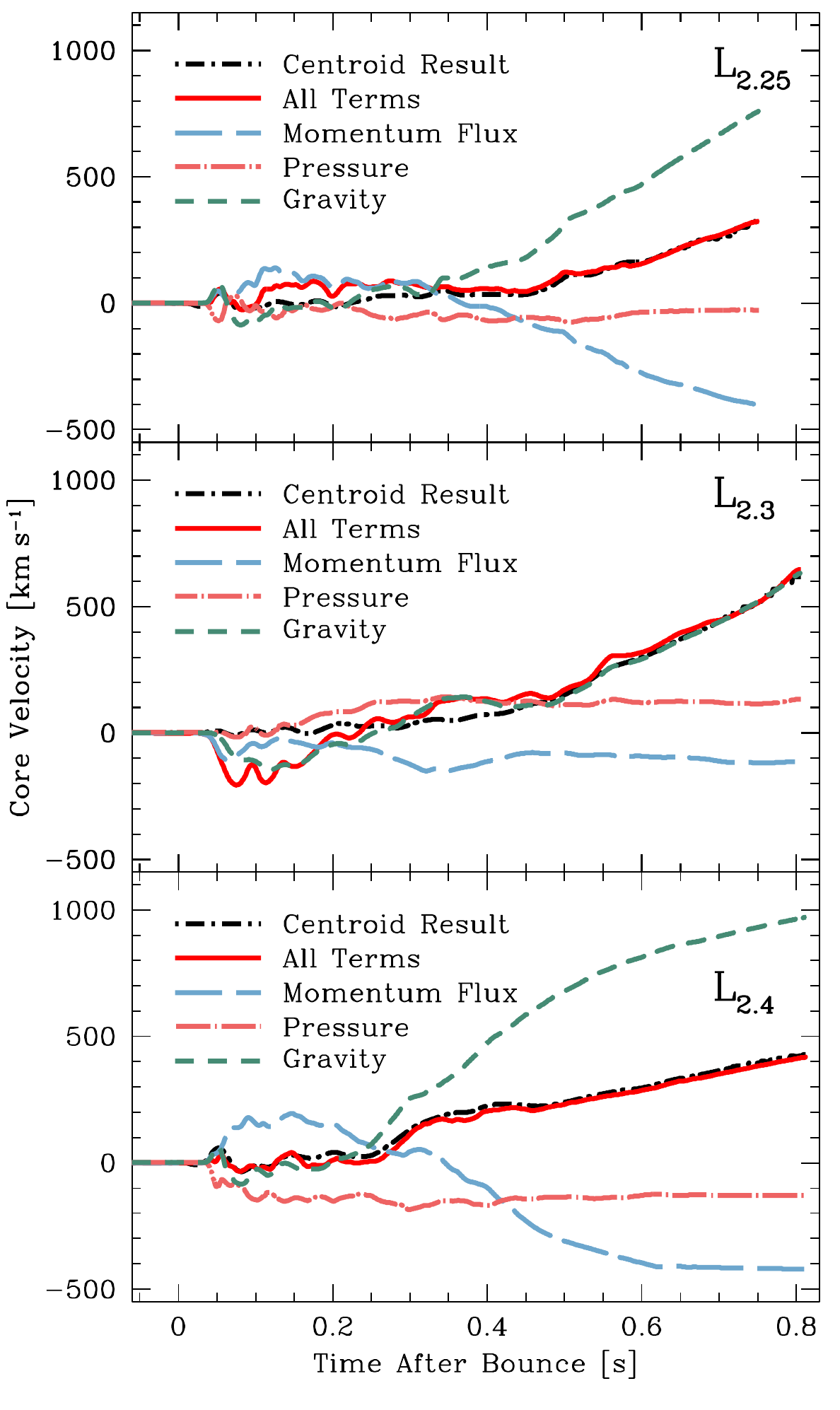}
\caption{Decomposed PNS kick velocities for models $\rm L_{2.25}$ (top
  panel), $\rm L_{2.3}$ (middle panel) and $\rm L_{2.4}$, obtained by
  integrating Eq.~\eqref{eq:acceleration} from bounce.  We have
  inverted the velocities for the $\rm L_{2.3}$ run.  The
  dash-dotted-black curve depicts the PNS velocity computed using the
  centroid of the density, while the solid red curve shows the
  contributions to the kick from momentum flux (long-dashed blue
  curve), gravity (dashed green curve) and pressure (dash-dot pink
  curve).  In each run, the PNS is still accelerating at more than 500
  km\,s$^{-2}$ at the end of the calculation, and this acceleration is
  dominated by the gravitational term. 
\label{fig:accel_3panel}}
\end{center}
\end{figure}

Using a fiducial radius of $r_{\rm c}=200$ km and integrating
Eq.~\ref{eq:acceleration} from core bounce, Figure
\ref{fig:accel_3panel} shows the decomposition of the PNS kick into
three components for models $\rm L_{2.25}$ (top panel), $\rm L_{2.3}$
(middle panel) and $\rm L_{2.4}$ (bottom panel).  Note that the
velocities have been reflected for model $\rm L_{2.3}$.  In each
panel, the dash-dotted black curve represents the smoothed centroid
velocity, while the solid red curve is the sum of the three terms in
Eq.~\eqref{eq:acceleration}.  The gravitational component
(short-dashed green curve)~dominates the late-time evolution in all
three simulations.  The anisotropic pressure term (dot-dashed pink
curve) flattens towards the end of each run as the PNS decouples from
the ejecta.  

Model $\rm L_{2.3}$ achieved the highest kick velocity in our suite of
simulations, more than 620 km\,s$^{-1}$ at $\sim$800 ms after bounce.
The middle panel of Figure~\ref{fig:accel_3panel} shows its velocity
evolution, decomposed using Eq.~\eqref{eq:acceleration} and inverted
to facilitate comparison with models $\rm L_{2.25}$ and $\rm L_{2.4}$.
Anisotropic pressure and momentum flux (dot-dashed pink and
long-dashed blue lines, respectively) contributed almost nothing to
the kick after $\sim$400 ms from core bounce.  Driven by the
gravitational term in Eq.~\eqref{eq:acceleration}, this model was
still accelerating at more than 1000 km\,s$^{-2}$ when the shock
reached the edge of the computational domain.  

While they did not achieve as large a PNS velocity as $L_{2.3}$,
models $\rm L_{2.25}$ and $\rm L_{2.4}$ were still accelerating at
$\sim$1000 and $\sim$600 km\,s$^{-2}$, respectively, at the end of our
calculations.  In both cases, and particularly for the $\rm L_{2.4}$
model, this acceleration was dominated by the gravitational term in
Eq.~\eqref{eq:acceleration}.  Figure \ref{fig:L_2p4_348} clearly shows
the PNS and ejecta in model $\rm L_{2.4}$ decoupling at $\sim$650 ms
after bounce (second panel from bottom), and having almost completely
decoupled by $\sim$800 ms after bounce.

The three models presented here comprise a representative sample of
our simulation results.  Table \ref{table1} presents additional
information on each of our runs, including the velocity, the explosion
energy $E_{\rm exp}$, and $\alpha$, a dimensionless measure of the
degree of asymmetry, at the end of the calculations.  The parameters
$\alpha$ and $E_{\rm exp}$ are defined by Eqs.~\eqref{eq:alpha} and
\eqref{eq:exp_ener} in the following section.  All of our calculations,
except for $\rm L_{2.1}$, ended when the shock approached a radius of
10,000 km; model $\rm L_{2.1}$ ended with $R_{\rm shock} = 3300$ km.
For a detailed discussion of the limitations of our approach, the
effect of fixing the neutrino luminosities and the reliability of the
late-time acceleration and explosion energies see Section
\ref{subsec:interp}.

\subsection{\label{subsec:asymmetry}Asymmetries in the Ejecta and Explosion Energies\protect}

The acceleration of the PNS depends on the dynamics of the
explosion and the evolution of the asymmetry of the shocked material.
This asymmetry may be quantified in various ways
\citep{Scheck:2006vn,Burrows:2007ly}; here, we adopt 
\begin{equation}
\alpha \equiv \langle v_z \rangle / \langle |v| \rangle ~,
\label{eq:alpha}
\end{equation}
where $\langle \rangle$
denotes a mass-weighted average over the post-shock region with $r >
100$ km (thereby excluding the PNS itself).  This choice is similar to
the $\alpha$ presented in \cite{Scheck:2006vn}.

\begin{table}
\caption{Parameters at the end of the simulation, when $R_{\rm shock}
  \sim 10$,000 km; $\alpha$ and $E_{\rm exp}$ are calculated using
  Eqs.~\eqref{eq:alpha} and \eqref{eq:exp_ener}, respectively.
  Model $\rm L_{2.1}$ ended with $R_{\rm shock} = 3300$ km.  Note that the PNS wind contributes $\sim$50\% of the explosion energies listed below. 
\label{table1}} 
\centering
  \begin{tabular}{@{}lccr@{}}
  \hline
 Model & $v_{\rm PNS}$ [km s$^{-1}$] & $E_{\rm exp}$ [$10^{51}$ erg] & $\alpha$ \\
\hline
${\rm L_{2.1}}$		& 	$-40$  	&  	0.29	&	$0.026$	\\
${\rm L_{2.15}}$	& 	$212$	&	0.69	&	$-0.25$		\\
${\rm L_{2.2}}$		&  	$-186$	&	0.89	&	$0.08$		\\
${\rm L_{2.25}}$	&	$315$	&	0.69	&	$-0.23$		\\
${\rm L_{2.3}}$		&	$-624$ 	&	1.13	&	$0.23$		\\
${\rm L_{2.35}}$	&	$194$	&	1.28	&	$-0.06$		\\
${\rm L_{2.4}}$		&	$431$	&    	1.23	&	$-0.15$		\\
${\rm L_{2.5}}$		&	$276$	&	0.99	&	$-0.10$	\\
\hline
\end{tabular}
\end{table}

Figure~\ref{fig:alpha} shows the evolution of $\alpha$ for models $\rm
L_{2.4}$ (solid blue curve) and $\rm L_{2.5}$ (dashed blue curve).
The solid red and dashed red curves depict the PNS recoil velocities
for the $\rm L_{2.4}$ and $\rm L_{2.5}$ models respectively.  Our
suite of simulations produced final values of $\alpha$ between $-0.25$
and 0.25 (see Table~\ref{table1}).  Since momentum is conserved,
larger values of $\alpha$ lead to larger PNS recoil velocities.  

Figure \ref{fig:shocks} shows the position of the shock at the end of
the calculation for five of our models; models with a negative kick
have been reflected.  While the shock asymmetry does correlate with
the kick velocity, the magnitude of the kick depends on the
distribution of matter behind the shock, which we paramtrize using
$\alpha$.

The third column of Table \ref{table1} shows the explosion energy at
the end of the simulation, defined as the total energy of all unbound
material on the grid,
\begin{equation}
\int_{E_{\rm tot}>0} \rho \left( u_{\rm int} + 
\frac{v^2}{2} - \frac{G}{r^2} M_{\rm enc} \right) d^3 \mathbf{x}~,
\label{eq:exp_ener}
\end{equation}
where $u_{\rm int}$ is the specific internal energy and $M_{\rm enc}$
is the mass interior to the fluid element.  At the end of our
calculations, the internal energy of shocked material dominates the
kinetic energy by a factor of $\sim$3--5, and the explosion energy is
still increasing due to sustained neutrino heating
(Eq.~\eqref{eq:heating}).  The bulk of this heating occurs primarily within the first few hundred kilometers of the PNS and is driving the late-time PNS wind.   

The internal energy of shocked material will ultimately be converted
into kinetic energy by $p\,dV$ work. In this limit, the PNS kick will
be a function of the anisotropy of the ejecta and $E_{\rm exp}$.
While the explosion will be nearly spherical in the outer envelope,
the anisotropy in the inner mass shells should freeze out at values
close to those indicated in Table 1.  This anisotropy in the inner
ejecta velocities, with the bulk of the ejecta traveling opposite to
the recoiling PNS, should be observable in the supernova remnant and 
is a specific prediction of our model.

\begin{figure}
\begin{center}
\includegraphics[width=8cm,angle=0,clip=true]{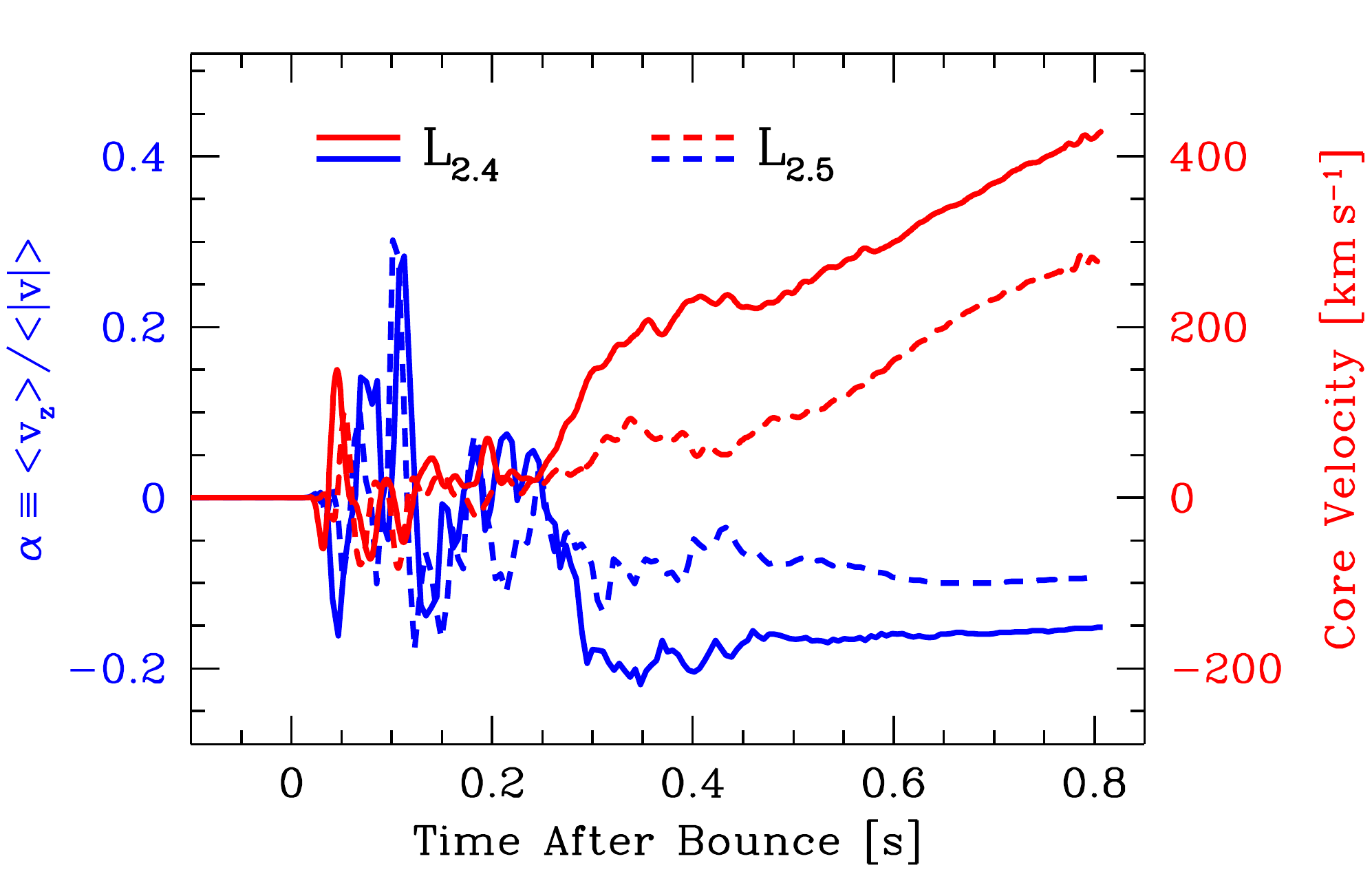}
\caption{The blue curves show the evolution of alpha
  (Eq.~\eqref{eq:alpha}) for models $\rm L_{2.4}$ (solid) and $\rm
  L_{2.5}$ (dashed), while the red curves show the core recoil
  velocity as a function of time.  The core velocity is always
  opposite the ejecta asymmetry due to conservation of momentum. 
\label{fig:alpha}}
\end{center}
\end{figure}

Figure~\ref{fig:misc} shows the total explosion energy (solid blue curve) as a function of time for model $\rm L_{2.4}$.  The explosion energy of material in the bottom half of the computational grid ($Z<0$) is depicted by the dot-dashed green curve, while the explosion energy of material in the top half of the computational grid ($Z>0$) is depicted by the dashed red curve.  Consistent with Fig.~\ref{fig:L_2p4_348}, this demontrates that model $\rm L_{2.4}$ explodes primarily in the $-Z$ direction.  The bottom panel of Fig.~\ref{fig:misc} shows the PNS mass as a function of time (solid blue curve) and the total mass of shocked, bound material exterior to the core (dot-dashed red curve).  By the end of model $\rm L_{2.4}$, the PNS mass is 1.45 $M_\odot$ while there is little shocked, bound matter outside the PNS itself.

\begin{figure}
\begin{center}
\includegraphics[width=8.5cm,angle=0,clip=true]{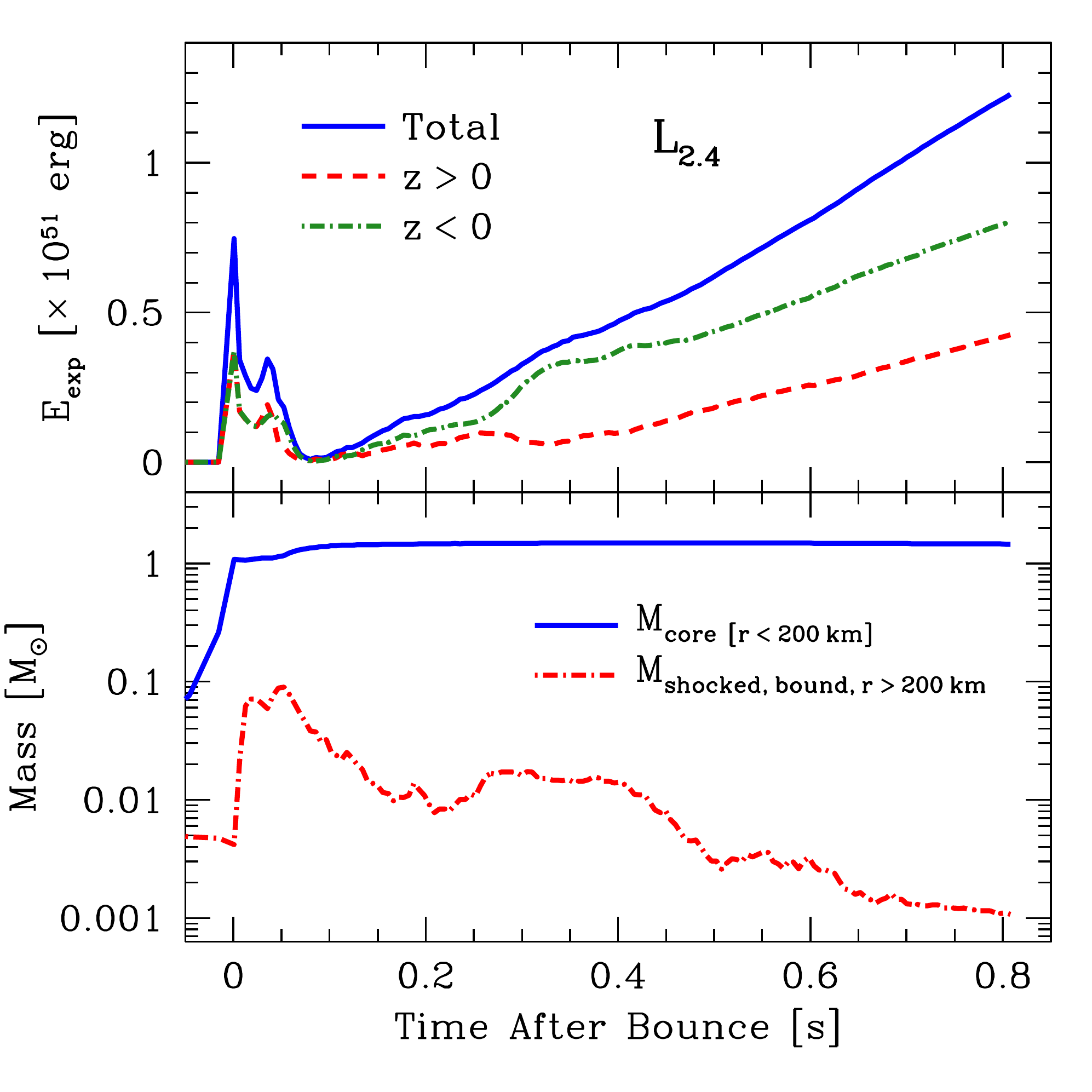}
\caption{The top panel shows the explosion energy (Eq.~\eqref{eq:exp_ener}) as a function of time for model $\rm L_{2.4}$.  The solid blue curve depicts the total explosion energy.  The explosion energy of material with $Z<0$ is shown by the dot-dashed green curve while the explosion energy of material with $Z>0$ is shown by the dashed red curve.  The bottom panel presents the PNS mass (solid blue curve) as a function of time; its final value is 1.45 $M_\odot$.  The dot-dashed red curve in the bottom panel shows the total shocked, bound mass exterior to the core as a function of time.  At the end of the simulation, only $\sim$$10^{-3}$ $M_\odot$ of shocked material outside the PNS remains bound.\label{fig:misc}}
\end{center}
\end{figure}

\begin{figure}
\begin{center}
\includegraphics[width=8.5cm,angle=0,clip=true]{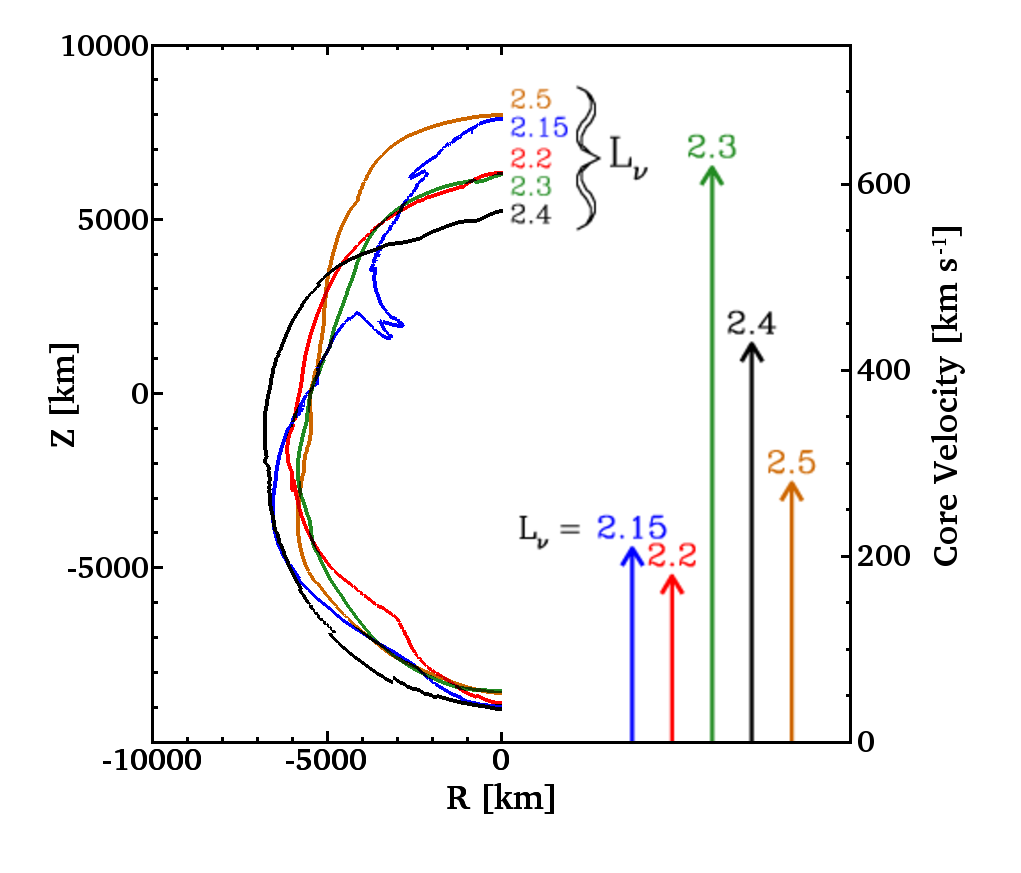}
\caption{The position of the outer edge of the gain region toward the end of the calculations.  Runs with negative PNS velocity have been reflected in this plot.  The arrows depict the core velocity for each run (right axis). \label{fig:shocks}}
\end{center}
\end{figure}

\subsection{\label{subsec:interp}Late-Time Evolution\protect}

Our simplified transport scheme allows us to perform multiple
calculations of the inner regions of a core-collapse supernova.  To
date, only one time-dependent, multi-group, flux-limited diffusion,
neutrino transport kick calculation, which includes the PNS on the
computational grid, exists in the literature
\citep{Nordhaus:2010kx}.  As computational methods and resources
improve, it will become possible to self-consistently connect the PNS
kick with the large-scale properties of the ejecta.  In this section,
we discuss the late-time evolution of our simulations and some of the
limitations of our approach.

As previously mentioned, the magnitude of the PNS kick will increase
with the degree of asymmetry in the ejecta and the explosion energy of
the supernova.  At the end of our calculations, the shocked matter's
internal energy exceeds its kinetic energy by a factor of $\sim$3--5.
Adiabatic expansion will convert this internal energy into kinetic
energy as the shock propagates through the stellar envelope.

Our constant driving $L_{\nu_e}$ also deposits energy into the
expanding ejecta, both by neutrino absorption in the gain region and
by driving a $\sim$$0.1$ $M_\odot$\,s$^{-1}$ wind from the surface of
the PNS.  This late-time wind contributes $\sim$50\% of the explosion energy at the end of each simulation and typically contains $\sim$$0.05$ $M_\odot$ of material.  While neutrino-driven winds from the PNS are expected \citep{Burrows:1995zr}, future improvements to this work should
include more sophisticated transport approaches which naturally
incorporate time-variable neutrino luminosities.  

We tested the effect of decaying neutrino driving luminosities by
restarting model $\rm L_{2.3}$ 400 ms after bounce with an
exponentially decreasing driving luminosity.  We used an exponential
decay timescale of 1 second, giving a $\sim$35\% reduction in
$L_{\nu_e}$ at the end of our calculation.  As a result of the lower
energy injection rate, the late-time PNS wind decreased by nearly a
factor of two, the PNS took longer to decouple from the post-shock
material, and the final PNS velocity was $\sim$25\% lower.  Still, the
PNS was accelerating gravitationally at $\sim$1000 km\,s$^{-1}$,
nearly as fast as in the model with a constant driving luminosity.

We also note that our calculations end when the shock reaches a fixed
radius of $\sim$10,000 km, rather than after a fixed amount of
post-bounce time.  The total amount of energy injected into our models
thus varies widely, making it difficult to connect the derived
explosion energies with our observed kick velocities.

\section{Comparison with Previous Numerical Work\label{4}}

Previous computational studies of pulsar recoil have employed various
simplifications and approximations to make the calculations tractable.
These approaches include excising the PNS from the computational
domain, starting the calculations $\sim$20 ms after bounce, and
employing simplified neutrino transport schemes
\citep{Scheck:2004rt,Scheck:2006vn,Wongwathanarat:2010yq}.  The
exclusion of the PNS from the domain is particularly useful,
as it avoids the severe Courant limitation imposed by resolving the
PNS.  The PNS is replaced by a rigid, spherical boundary, which
contracts according to a prescription from a detailed spherically-symmetric-collapse calculation.  This approach is designed to mimic the settling of material as the PNS cools.

By using all three of these simplifications, previous studies have
been able to track the shock to large distances ($>10^4$ km) and late
times ($>$1 s)
\citep{Scheck:2004rt,Scheck:2006vn,Wongwathanarat:2010yq}.  While
useful for calculating long-term evolution, this approach requires
inferring recoil through a rigid boundary of infinite inertial mass.
Furthermore, this approach neglects effects resulting from the
displacement of the PNS relative to the surrounding matter.  To
compensate for this effect, these authors have added a
kick to the gas which mimics movement of the PNS.  The physical
fidelity of such approximations has been verified by self-consistent
calculations such as those presented in \citep{Nordhaus:2010kx} and in
this work.

Recently, \cite{Nordhaus:2010kx} presented the first axisymmetric,
multi-group, flux-limited diffusion neutrino transport calculation of
recoil in which core collapse lead to significant acceleration of a
fully-formed PNS.  The authors used the multi-group, arbitrary
Lagrangian-Eulerian (ALE), radiation-hydrodynamics code {\sc
VULCAN/2D} \citep{Livne:1993zr}.  The calculation employed multi-group
flux-limited diffusion neutrino transport \citep{Livne:2004zr},
supplemented the neutrino luminosity by an additional
$L_{\nu_e}=L_{\bar{\nu}_e}=2 \times 10^{52}$ erg\,s$^{-1}$, and used
the same 15-$M_\odot$ progenitor as this work.  During the induced,
neutrino-driven explosion, a $\sim$10\% anisotropy in the ejecta led
to a PNS recoil velocity of $\sim$150 km\,s$^{-1}$ at the end of the
calculation, when the shock reached a radius of $\sim$5,000 km.  Such
a result in terms of PNS velocity and ejecta asymmetry compares
favorably with model $\rm L_{2.2}$ presented in this work (see
Table~\ref{table1}) and the results of \cite{Scheck:2006vn}.

In general, given the different computational techniques and the use
of three different codes, the agreement between our results and those
of \citeauthor{Scheck:2006vn} and \cite{Nordhaus:2010kx} is
gratifying.  Our detailed calculations of the first second of
post-bounce evolution produce high-velocity recoils comparable with
those in \cite{Scheck:2004rt,Scheck:2006vn} and \cite{Nordhaus:2010kx}
while following the evolution of the PNS itself.  Taken together,
these studies strongly suggest that asymmetric core-collapse
supernovae naturally lead to acceleration of the PNS and are capable
of birthing the highest velocity pulsars.

While axisymmery would restrict
core motion to the $Z$-axis, 3D computations impose no such constraint
and allow one to measure the PNS spin in addition to recoil.  Recently, \cite{Wongwathanarat:2010yq} presented the first 3D
excised-core calculations of recoil.  High PNS velocities were achieved for rotating and non-rotating progenitors, providing further evidence that PNSs are naturally accelerated during core collapse.  However, in the case of pulsar spins,
\cite{Rantsiou:2011lr} showed that excising the PNS from the computational domain can lead to qualitatively
different results in the spin rates.  As such, future 3D calculations which include the
PNS should be performed and differences between kicks from
different progenitor models (rotating and non-rotating) should be investigated.

\section{Conclusions\label{5}}
We have carried out a suite of axisymmetric simulations of the
collapse of a massive star's core with the AMR, radiation-hydrodynamic
code {\sc CASTRO}.  For each calculation, we follow the core collapse,
PNS formation, ensuing neutrino-driven supernova explosion and PNS
recoil.  By incorporating the effects of neutrino heating and cooling
in place of more detailed and computationally expensive neutrino
transport, we are able to perform multiple calculations that
simultaneously follow the evolution of the PNS and the global
explosion for $\sim$1 second and to distances of $\sim$10,000 km.

The PNSs in our simulations achieved recoil velocities comparable to
the those of observed pulsars.  After $\sim$1 second of post-bounce
evolution, the highest PNS velocity obtained was 620 km\,s$^{-1}$
(model $\rm L_{2.3}$).  After $\sim$0.6 seconds of post-bounce
evolution, this acceleration was supplied primarily by the
gravitational pull of slow-moving ejecta in front of the PNS.  This
gravitational effect dominates the late-time PNS acceleration in all
of our calculations.  While our PNSs have started to decouple from the
surrounding fluid (see Fig.~\ref{fig:accel_3panel}), the substantial
and ongoing gravitational acceleration suggests that higher velocities
may ultimately be achievable.

Our results suggest that hydrodynamic recoil during an asymmetric
supernova explosion provides a natural explanation for pulsar kicks.
After the bounce shock stalls, hydrodynamic instabilities deform the
shocked material and ensure that the ensuing explosion is asymmetric.
Recoil during the supernova explosion and gravitational interaction
with the expanding ejecta subsequently accelerate the PNS to high
velocities.  The results presented in this work are consistent with the findings of \cite{Nordhaus:2010kx} and previous axisymmetric simulations that excised the PNS from the computational domain \citep{Scheck:2004rt,Scheck:2006vn}.  Taken together, these studies strongly suggest that generic core-collapse supernovae can accelerate neutron stars to the high velocities observed in the pulsar population.  Additionally, these studies demonstrate that hydrodynamic processes, and not asymmetric neutrino emission, are responsible for this acceleration \citep{Scheck:2006vn,Burrows:2007ly,Nordhaus:2010kx}.  In fact, recent simulations of neutron star kicks in three dimensions suggest that velocities comparable to those from axisymmetric calculations are achievable \citep{Wongwathanarat:2010yq}.

In this work, we have provided substantial numerical support to the
hydrodynamic mechanism of pulsar kicks. Recoil due to a
neutrino-driven, core-collapse supernova explosion provides a natural
explanation for pulsar kicks without appealing to more exotic
scenarios.  As computational methods and resources improve,
self-consistent three-dimensional studies will enable a full
comparison of theoretical models to observed distributions of pulsar
kicks and spins.  

\section*{Acknowledgements}
The authors thank Noam Soker, Thomas Janka, Annop Wongwathanarat and Ewald Mueller for comments which lead to an improved manuscript.  JN is supported by an NSF Astronomy and Astrophysics Postdoctoral Fellowship under award AST-1102738 and by NASA HST grant AR-12146.04-A.  TBD is supported by an NSF Graduate Research Fellowship under grant number DGE-0646086.  AB is supported by the Scientific Discovery through Advanced Computing (SciDAC) program of the DOE, under grant DE-FG02-08ER41544, the NSF under the subaward ND201387 to the Joint Institute for Nuclear Astrophysics (JINA, NSF PHY-0822648), and the NSF PetaApps program, under award OCI-0905046 via a subaward 44592 from Louisiana State University to Princeton University. Work at LBNL was supported in part by the SciDAC program under contract DE-FC02-06ER41438.  AA is supported by the Office of High Energy Physics and the Office of Mathematics, Information, and Computational Sciences as part of the SciDAC Program under the U.S. Department of Energy under contract No.~DE-AC02-05CH11231.The authors thank the members of the Center for Computational Sciences and Engineering (CCSE) at LBNL for their invaluable support for CASTRO.

Support for HPC storage and resources was provided by the National Energy Research Scientific Computing Center (NERSC), which is supported by the Office of Science of the US Department of Energy under contract DE-AC03-76SF00098; by NICS, on the Kraken supercomputer, provided by the National Science Foundation through the TeraGrid Advanced Support Program under grant TG-AST100001; by the TIGRESS High Performance Computing and Visualization Center at Princeton University, which is jointly supported by the Princeton Institute for Computational Science and Engineering (PICSciE) and the Princeton University Office of Information Technology.

\bibliography{NSkicks}
\bibliographystyle{astron}

\end{twocolumn}

\end{document}